\documentclass{ws-procs975x65}

\begin{document}


\wstoc{Solution generating theorems: perfect fluid spheres and the TOV equation}{P. Boonserm, M. Visser, S. Weinfurtner}

\title{SOLUTION GENERATING THEOREMS: PERFECT FLUID SPHERES AND THE TOV EQUATION\footnote{This research was supported by the Marsden Fund administered by the Royal Society of New Zealand. In addition, PB was supported by a Royal Thai Scholarship and a Victoria University Small Research Grant. SW was supported by the Marsden Fund, by a Victoria University PhD Completion Scholarship, and a Victoria University Small Research Grant.}}

\author{Petarpa Boonserm, Matt Visser, and Silke Weinfurtner}
\address{School of Mathematics, Statistics, and Computer Science,\\ 
Victoria University of Wellington, PO Box 600, Wellington, New Zealand\\
\email{Petarpa.Boonserm@mcs.vuw.ac.nz, matt.visser@mcs.vuw.ac.nz, silke.weinfurtner@mcs.vuw.ac.nz}}

\begin{abstract}
We report several new transformation theorems that map perfect fluid spheres into perfect fluid spheres. In addition, we report new ``solution generating'' theorems for the TOV, whereby any given solution can be ``deformed'' to a new solution.
\end{abstract}

\bodymatter

\section{Introduction}\label{intro}
Perfect fluid spheres, either Newtonian or relativistic, are the first approximation in developing realistic stellar models~\cite{Delgaty, Skea, exact}.
For our current purposes, the central idea is to start solely with spherical symmetry, which implies that in orthonormal components the stress energy tensor takes the form:
\begin{equation}
T_{\hat a\hat b} = \left[ \begin{array}{cccc}
\rho&0&0&0\\ 0&p_r&0&0\\ 0&0&p_t&0\\ 0&0&0&p_t \end{array}\right],
\end{equation}
and then use the perfect fluid constraint $p_r=p_t$. This simply makes the radial pressure equal to the transverse pressure.
By using the Einstein equations, plus spherical symmetry, the equality  $p_r=p_t$ for the pressures becomes the statement
$
G_{\hat\theta\hat\theta} = G_{\hat r\hat r} = G_{\hat\phi\hat\phi}
$.
\section{Solution generating theorems}
Start with some static spherically symmetric geometry in Schwarzschild (curvature) coordinates
\begin{equation} \label{line_element_1}
ds^2 = - \zeta(r)^2 \; dt^2 + {dr^2\over B(r)} + r^2 \;d\Omega^2,
\end{equation}
and assume it represents a perfect fluid sphere. Setting $G_{\hat r\hat r} = G_{\hat\theta\hat\theta}$ supplies us with an ODE
\begin{equation} 
\label{ode_for_B}
[r(r\zeta)']B'+[2r^2\zeta''-2(r\zeta)']B + 2\zeta=0 \, ,
\end{equation}
Solving for $B(r)$ in terms of $\zeta(r)$ is the basis of the analyses in references \cite{Lake,Martin}. On the other hand,
we can also re-group this same equation as
\begin{equation}    
\label{ode_for_zeta}
2 r^2 B \zeta'' + (r^2 B'-2rB) \zeta' +(r B'-2B+2)\zeta=0 \,,
\end{equation}
which is a linear homogeneous 2nd order ODE for $\zeta(r)$. Suppose we start with the specific geometry defined by
\begin{equation}
ds^2 = - \zeta_0(r)^2 \; dt^2 + {dr^2\over B_0(r)} + r^2 d\Omega^2
\end{equation}
and assume it represents a perfect fluid sphere.
We will show how to ``deform'' this solution by applying four different transformation theorems on $\left\{ \zeta_0 , B_0  \right\}$.
\subsection{Four theorems}
The first theorem we present is a variant of a result  first explicitly published in reference \cite{Martin}.
\paragraph{Theorem 1} 
Suppose $\{ \zeta_0(r), B_0(r) \}$ represents a perfect fluid sphere.
Define
\begin{equation} \label{Theorem1_m_1}
\Delta_0(r)  =
 \left({ \zeta_0(r)\over  \zeta_0(r) + r  \;\zeta'_0(r)}\right)^2 \; r^2 \; 
\exp\left\{ 2 \int {\zeta'_0(r)\over  \zeta_0(r)} \; 
  { \zeta_0(r)- r\; \zeta'_0(r)\over  \zeta_0(r) + r  \;\zeta'_0(r)} \; d r\right\}.
\end{equation}
Then for all $\lambda$, the geometry defined by holding $\zeta_0(r)$ fixed and
setting
\begin{equation}
ds^2 = - \zeta_0(r)^2 \; dt^2 + {dr^2\over B_0(r)+\lambda\; \Delta_0(r) }
+ r^2 d\Omega^2
\end{equation}
is also a perfect fluid sphere.
\paragraph{Theorem 2} 
Let $\{\zeta_0,B_0\}$ describe a perfect fluid sphere.
Define
\begin{equation}
Z_0(r) = \sigma +\epsilon \int {r \; dr\over  \zeta_0(r)^2\; \sqrt{B_0(r)} }.
\end{equation}
Then for all $\sigma$ and $\epsilon$, the geometry defined by holding $B_0(r)$ fixed and
setting
\begin{equation}
ds^2 = - \zeta_0(r)^2 \; Z_0(r)^2
\; dt^2 + {dr^2\over B_0(r)} + r^2 d\Omega^2
\end{equation}
is also a perfect fluid sphere.


Having now found the first and second generating theorems it is possible to define two new theorems by composing them.
Take a perfect fluid sphere solution $\left\{ \zeta_0, B_0  \right\}$. Applying Theorem 1 onto it gives
us a new perfect fluid sphere $\left\{ \zeta_0, B_1 \right\}$. The new $B_1$ is given in
equation (\ref{Theorem1_m_1}). If we now continue by applying Theorem 2, again we get a
new solution $\{ \tilde\zeta, B_1 \}$, where $\tilde{\zeta}$ now depends on the new $B_1$. For more details regarding Theorem 3 and Theorem 4 see reference \cite{Petarpa}.
\section{Solution generating theorems for the TOV equation}
The Tolman--Oppenheimer--Volkov [TOV] equation constrains the internal structure of general relativistic static perfect fluid spheres \cite{Petarpa1}. In this analysis the pressure and density are primary and the spacetime geometry is secondary.
Using standard results (see the explicit discussion in reference~\cite{Petarpa1}) it is relatively simple to present
the following:
\paragraph{Theorem {\bf P1}}
We derived Theorem {\bf P1} by looking for changes in the pressure profile with $m_0$ fixed. This theorem can also be viewed as a consequence of Theorem 2. The key difference now is that we have an explicit statement directly in terms of the shift in the pressure profile \cite{Petarpa1}.
\paragraph{Theorem {\bf P2}}
A second theorem can be obtained by looking for \emph{correlated}
changes in the mass and pressure profiles. In addition, we can also view this Theorem {\bf P2} as a consequence of Theorem 1 . The key difference now is that we have an explicit statement directly in terms of the shift in the pressure profile \cite{Petarpa1}.
\section{Discussion}

Using Schwarzschild coordinates we have developed two fundamental transformation theorems that map perfect fluid spheres into perfect fluid spheres. Moreover, we have also established two additional transformation theorems by composing the first and second generating theorems.
Furthermore, we have also developed two ``physically clean'' solution-generating theorems for the TOV equation --- where by
``physically clean'' we mean that it is relatively easy to understand
what happens to the pressure and density profiles, especially in the
vicinity of the stellar core.


\begin{thebibliography}{99}

\bibitem{Delgaty}
M.~S.~R.~Delgaty and K.~Lake,
``Physical acceptability of isolated, static, spherically symmetric,  perfect
fluid solutions of Einstein's equations,''
Comput.\ Phys.\ Commun.\  {\bf 115} (1998) 395
[arXiv:gr-qc/9809013].

\bibitem{Skea}
M.~R.~Finch and J.~E.~F.~Skea,
``A review of the relativistic static fluid sphere'',
1998, unpublished. {\sf http://www.dft.if.uerj.br/usuarios/JimSkea/papers/pfrev.ps}

\bibitem{exact}
H. Stephani, D. Kramer, M. MacCallum, C. Hoenselaers, and E. Herlt,
\emph{Exact solutions of Einstein's field equations},
  (Cambridge University Press, 2003).
  
 

\bibitem{Lake}
K.~Lake,
``All static spherically symmetric perfect fluid solutions of Einstein's
Equations,''
Phys.\ Rev.\ D {\bf 67} (2003) 104015
[arXiv:gr-qc/0209104].


\bibitem{Martin}
D.~Martin and M.~Visser,
``Algorithmic construction of static perfect fluid spheres,''
Phys.\ Rev.\ D {\bf 69} (2004) 104028
[arXiv:gr-qc/0306109].

\bibitem{Petarpa}
P.~Boonserm, M.~Visser, and S.~Weinfurtner
``Generating perfect fluid spheres in general relativity,'' 
Phys. \ Rev. \ D {\bf 71} (2005) 124037.
[arXiv:gr-qc/0503007].

\bibitem{Petarpa1}
P.~Boonserm, M.~Visser, and S.~Weinfurtner
``Solution generating theorems for the TOV equation,'' 
[arXiv:gr-qc/0607001].
  
  















%



\end{thebibliography}
\end{document}